\documentclass[preprint,aps]{revtex4}
\usepackage{amsfonts}
\usepackage{graphicx}
\usepackage{bm}

\begin{document}

\title{Topological invariance and global Berry phase in non-Hermitian systems}
\author{Shi-Dong Liang}
\altaffiliation{Corresponding author: Email: stslsd@mail.sysu.edu.cn}
\author{Guang-Yao Huang}
\affiliation{State Key Laboratory of Optoelectronic Material and Technology and,
School of Physics and Engineering, Sun Yat-Sen University, Guangzhou, 510275,
People's Republic of China}
\date{\today }

\begin{abstract}
Studying the topological invariance and Berry phase in non-Hermitian systems, we give the basic properties of the complex Berry phase and generalize the global Berry phases $Q$ to identify the topological invariance to non-Hermitian models. We find that $Q$ can identify a topological invariance in two kinds of non-Hermitian models, two-level non-Hermitian Hamiltonian and bipartite dissipative model.
For the bipartite dissipative model, the abrupt change of the Berry phase in the parameter space reveals quantum phase transition and relates to the exceptional points. These results give the basic relationships between the Berry phase, quantum and topological phase transitions.
\end{abstract}

\maketitle



\section{Introduction}
Quantum state storage and transfer are central issues of quantum technology.
The basic idea is to find an efficient way to generate robust quantum states.\cite{Prinz} Several schemes of quantum devices, such as quantum adiabatic pumps\cite{Zhou} and geometric quantum computation, exhibit excellent features.\cite{Paolo,Zhu}The geometric phase plays a crucial role in realizing robust quantum states and detecting quantum phase transition (QPT).\cite{Carollo,Sachdev}
The geometric phase generalized to non-Hermitian systems provides a geometrical description of the quantum evolution of non-Hermitian systems,\cite{Hossein} and give the relationship between geometric phase and QPT.\cite{Nesterov}
For the non-Hermitian quantum walk, it has been found that the topological invariance relies on the ratio of the hopping amplitude between A and B sites.\cite{Rudner} The topological invariance provides some hints to realize robust quantum states and opens some fundamental issues.\cite{Nussinov} The challenging problem is how to give a unified definition of the topological invariance for different quantum systems.
In general, the topological phase transition (TPT) is characterized by topological index instead of symmetry breaking, such as winding number and Chern number.\cite{Wen} For example, the integer conductance plateau in the integer quantum Hall effect is related to the topological quantity Chern class and geometric phase,\cite{Shapere} which reveals the relationship between TPT and geometric phase.

However, there has been no a general method to define the topological invariance
for different quantum systems, particularly for non-Hermitian systems.
Actually, how to give a unified definition of topological invariance and how to characterize TPT for a general system are still challenging problems even though there have been some schemes for some specific systems.\cite{Nussinov,Wen,Kitaev} Hence, it is valuable for any effort to construct a new paradigm to
describe the topological invariance and TPT for both Hermitian and non-Hermitian systems.

In this paper, we will study the basic properties of the complex Berry phase in quantum evolution of non-Hermitian models. We propose the global Berry phase $Q$ of all states to identify the topological invariance, which is used to two kinds of non-Hermitian models. One is a general $2\times 2$ non-Hermitian model, the other is a bipartite dissipative model. They have not same mathematical structures. We find that $Q$ can be a topological index to identify the topological invariance in these two models. We will also give the phase diagram of the complex geometric phase in the parameter space that indicates the relationships between the topological invariance, quantum phase transition and the complex Berry phase.

\section{Berry phase in quantum evolution of non-Hermitian systems}
We consider a general quantum system described by the parameter-dependent Hamiltonian
$H(\mathbf{\alpha})$, where the parameters $\mathbf{\alpha}=(\alpha_{1},\alpha_{2},...\alpha_{m})$ are functions of time, which lead to an implicit time dependence of $H(\mathbf{\alpha})$ and describes the state evolution.
For a general non-Hermitian system, $H(\mathbf{\alpha})\neq H^{\dag}(\mathbf{\alpha})$. The state vectors $|\Psi(\mathbf{\alpha})\rangle$ and $|\Lambda(\mathbf{\alpha})\rangle$ within Hilbert space, $|\Psi(\mathbf{\alpha})\rangle \in \mathcal{H}$ and its dual Hilbert space $|\Lambda(\mathbf{\alpha})\rangle \in \mathcal{H^{\dagger}}$ satisfy the Schr\"{o}dinger equation,\cite{Keck}
\begin{equation}
\left\{
\begin{array}{c}
i\hbar \partial _{t}|\Psi(\mathbf{\alpha}) \rangle =H(\mathbf{\alpha})|\Psi(\mathbf{\alpha})\rangle  \\
i\hbar \partial _{t}|\Lambda(\mathbf{\alpha})\rangle =H^{\dagger}(\mathbf{\alpha})|\Lambda(\mathbf{\alpha})\rangle
\end{array}
\right.
\label{Seq}
\end{equation}

In the adiabatic approximation, the state evolution can be expanded into the instantaneous right eigen states
\begin{equation}
|\Psi({\mathbf{\alpha}})\rangle =\sum_{\mu}c_{\mu}(t)|\phi _{\mu}({\mathbf{\alpha}})\rangle,
\label{psi}
\end{equation}
where $|\phi _{\mu}(\mathbf{\alpha})\rangle$ satisfies the instantaneous eigen equation $H(\mathbf{\alpha})|\phi _{\mu}(\mathbf{\alpha})\rangle=E_{\mu}(\mathbf{\alpha})|\phi _{\mu}(\mathbf{\alpha})\rangle $.

Neglecting the off-diagonal terms in the adiabatic approximation, the coefficient can be expressed as\cite{Keck}
\begin{equation}
c_{\mu}(t)=c_{\mu}(0) e^{-\frac{i}{\hbar}\int_{0}^{t}E_{\mu}(t')dt'}
e^{i\int_{0}^{t}\langle\lambda_{\mu}({\mathbf{\alpha}})|\nabla_{\alpha}|\phi _{\mu}({\mathbf{\alpha}})\rangle\cdot d{\mathbf{\alpha}}},
\label{ct}
\end{equation}
where $\langle\lambda_{\mu}({\mathbf{\alpha}})|$ is the corresponding eigen state of $|\phi _{\mu}(\mathbf{\alpha})\rangle$ in the dual space.
The parameters $\alpha$ are varied in a cyclic way and the initial state
of the system stays at an instantaneous eigen state
$|\Psi(0)\rangle=|\phi_{\mu}(\alpha(0))\rangle$, the state evolution follows\cite{Keck}
\begin{equation}
|\Psi(T)\rangle=e^{i\left(\gamma^{D}_{\mu}(T)+\gamma^{G}_{\mu}(T)\right)}|\Psi(0)\rangle
\label{psi2}
\end{equation}
where $\gamma^{D}_{\mu}=-\frac{1}{\hbar}\int_{0}^{T}E_{\mu}(t)dt$ is
the complex dynamical phase.
$T$ is the periodicity of time evolution such that $\alpha(T)=\alpha(0)$.
$\gamma^{D}_{\mu}$ can be written as $\gamma^{D}_{\mu}\equiv \gamma^{d}_{\mu}+i\xi^{d}_{\mu}$, where the real part $\gamma^{d}_{\mu}$ is the dynamical phase, and the imaginary part $\xi^{d}_{\mu}$ is the dissipative effect induced by the energy shift for non-Hermitian systems. The complex Berry phase in the cyclic evolution is
\begin{equation}
\gamma^{B}_{\mu}=\oint_{C}A^{\mu}
\end{equation}
where $A^{\mu}=i\langle\lambda_{\mu}({\mathbf{\alpha}})|d|\phi _{\mu}({\mathbf{\alpha}})\rangle$ is the Berry potential (connection), where $d$ is the exterior derivative. Similarly, $\gamma^{B}_{\mu}\equiv \gamma^{b}_{\mu}+i\xi^{b}_{\mu}$, where
the real part $\gamma^{b}_{\mu}$ is the Berry phase, and the imaginary part
$\xi^{b}_{\mu}$ is the dissipative effect induced by the geometric potential.

For non-Hermitian systems, the dissipative effect originates from $\xi^{d}_{\mu}$ and $\xi^{b}_{\mu}$, where $\xi^{d}_{\mu}$ depends on the eigenvalues of $H$ and describes the dynamical dissipation. $\xi^{b}_{\mu}$ depends on the geometry of the evolution path. We refer them to the geometrical dissipative effect.

We may define a non-Abelian Berry connection (gauge potential one form) by \cite{Hirano}
\begin{eqnarray}
A & = & A_{k} d \alpha^{k} =i\Lambda^{\dagger}d \Psi \\\nonumber
&=&
i \left(\begin{array}{ccc}
\langle\lambda_{1}|d\psi_{1}\rangle & \cdots &\langle\lambda_{1}|d\psi_{M}\rangle \\
\vdots & \ddots & \vdots \\\nonumber
\langle\lambda_{M}|d\psi_{1}\rangle & \cdots &\langle\lambda_{M}|d\psi_{M}\rangle \\\nonumber
\end{array}
\right),
\end{eqnarray}
where
\begin{equation}
A_{k}=
i \left(\begin{array}{ccc}
\langle\lambda_{1}|\partial_{k}\psi_{1}\rangle & \cdots &\langle\lambda_{1}|\partial_{k}\psi_{M}\rangle \\
\vdots & \ddots & \vdots \\\nonumber
\langle\lambda_{M}|\partial_{k}\psi_{1}\rangle & ...&\langle\lambda_{M}|\partial_{k}\psi_{M}\rangle \\\nonumber
\end{array}
\right).
\end{equation}

It should be remarked that the non-Abelian Berry connection involves all eigen states of the system such that it can describe the global properties of the system.
We may define the global Berry phase of the system,\cite{Zhou}
\begin {equation}
Q=\frac{1}{2\pi}\oint Tr A.
\end{equation}
We will demonstrate that the global Berry phase $Q$  can be used to identify the topological invariance for Hermitian and two kinds of  non-Hermitian systems. We call $Q$ as the global Berry phase or the ground-state Berry phase. Similarly to the Hermitian systems, we can obtain a few of basic properties on the complex Berry phase.

{\bf Claim 2.1}: For the gauge transformation $|\phi_{\mu}'\rangle =e^{-if(\alpha)}|\phi_{\mu}\rangle$
where $e^{-if(\alpha)}$ is a single-value function modulo $2\pi$, namely
$f\left[\alpha(t)+\alpha(t+T)\right]=f[\alpha(t)]+2n\pi$, where $n$ is integer,\\
(a) the Berry potential is $A'_{\nu}=A_{\nu}+df$; \\
(b) the complex Berry phase is invariant mod $2\pi$, $\gamma'^{G}_{\nu}=\gamma^{G}_{\nu}+2n_{\nu}\pi$;\\
(c) $Q'=Q+\sum_{\nu}n_{\nu}$,
where $n_{\nu}$ is integer corresponding to the state $\nu$.\\
{\bf Proof}: (a) $A'_{\nu}=i\langle \lambda_{\nu}(\alpha)|e^{if(\alpha)}de^{-if(\alpha)}|\phi_{\nu}(\alpha)\rangle=
i\langle\lambda_{\nu}(\alpha)|d|\phi_{\nu}(\alpha)\rangle+i\langle \lambda_{\nu}(\alpha)|\phi_{\nu}(\alpha)\rangle e^{if(\alpha)}de^{-if(\alpha)}=A_{\nu}+df$;
(b) notice that $\oint df=2n_{\nu}\pi$, we have $\gamma'^{G}_{\nu}=\oint A'_{\nu}=\gamma^{G}_{\nu}+2n_{\nu}\pi$;
(c) Straightforwardly since $Tr A'=Tr A+Tr(df)$, and $\oint Trdf=Tr\oint df=\sum_{\mu}n_{\mu}$, hence $Q'=\frac{1}{2\pi}\oint TrA'=Q+\sum_{\mu}n_{\mu}$.$\Box$

It can be seen that $\gamma^{G}_{\nu}$ and $Q$ are gauge invariant modulo $2\pi$. They will be used to describe quantum phase transition and topological invariance of systems.

\section{Topological invariance and quantum phase transition in Non-Hermitian Models}
\subsection{Two-level non-Hermitian model}
Let us consider a $2\times2$ non-Hermitian Hamiltonian,
\begin{equation}
H=H_{hermi}+H_{nonhermi}
\label{Hn1}
\end{equation}
where the Hermitian part of the Hamiltonian is given as
\begin{equation}
H_{hermi}=\mathbf{h}\cdot\mathbf{\sigma},
\end{equation}
where  $\mathbf{h}(\alpha)=(h_{x}\sin\theta_{\alpha}\cos\varphi_{\alpha},h_{y}\sin\theta_{\alpha}\sin\varphi_{\alpha},
h_{z}\cos\theta_{\alpha})$ describes the Hermitian properties. Thus,
\begin{equation}
H_{hermi}=\left[\begin{array}{cc}
h_{z}\cos\theta_{\alpha} & (h_{x}\cos\varphi_{\alpha}-ih_{y}\sin\varphi_{\alpha})\sin\theta_{\alpha}\\
(h_{x}\cos\varphi_{\alpha}+ih_{y}\sin\varphi_{\alpha})\sin\theta_{\alpha} & -h_{z}\cos\theta_{\alpha}
\end{array}\right]
\end{equation}
The non-Hermitian part is
\begin{equation}
H_{nonhermi}=\mathbf{\Delta}\cdot\mathbf{n}(\alpha),
\end{equation}
where
\begin{equation}
\mathbf{\Delta}=
\left(\begin{array}{cc}
0 & \Delta_{x} \\
-\Delta_{x} & 0
\end{array}\right)\mathbf{i}+
\left(\begin{array}{cc}
0 & -i\Delta_{y} \\
-i\Delta_{y} & 0
\end{array}\right)\mathbf{j}+
\left(\begin{array}{cc}
i\Delta_{z} & 0 \\
0 &-i\Delta_{z}
\end{array}\right)\mathbf{k}
\end{equation}
describes the non-Hermitian properties, and $\mathbf{n}(\alpha)=(\sin\theta_{\alpha}\cos\varphi_{\alpha},\sin\theta_{\alpha}\sin\varphi_{\alpha},\cos\theta_{\alpha})$
is a unit vector of Bloch sphere, we have
\begin{equation}
H_{nonhermi}= \left[\begin{array}{cc}
i\Delta_{z}\cos\theta_{\alpha} & (\Delta_{x}\cos\varphi_{\alpha}-i\Delta_{y}\sin\varphi_{\alpha})\sin\theta_{\alpha}\\
(-\Delta_{x}\cos\varphi_{\alpha}-i\Delta_{y}\sin\varphi_{\alpha})\sin\theta_{\alpha} & -i\Delta_{z}\cos\theta_{\alpha}
\end{array}\right],
\end{equation}
where $\alpha$ is the evolution parameter. The total Hamiltonian is rewritten as
\begin{equation}
H=\left[
\begin{array}{cc}
Z\cos\theta_{\alpha} & r_{\alpha}^{(+)}e^{i\nu_{\alpha,1}}\sin\theta_{\alpha} \\
r_{\alpha}^{(-)}e^{i\nu_{\alpha,2}}\sin\theta_{\alpha} & -Z\cos\theta_{\alpha}
\end{array}
\right];
\label{nonH2}
\end{equation}
where $Z=h_z+i\Delta_{z}$;
$r_{\alpha}^{(\pm)}=\sqrt{(h_x\pm\Delta_x)^{2}\cos^{2}\varphi_{\alpha}+(h_y\pm\Delta_y)^{2}\sin^{2}\varphi_{\alpha}}$;
$\nu_{\alpha,1}=\arctan\left[-\frac{h_y+\Delta_y}{h_x+\Delta_x}\tan\varphi_{\alpha}\right]$;
$\nu_{\alpha,2}=\arctan\left[\frac{h_y-\Delta_y}{h_x-\Delta_x}\tan\varphi_{\alpha}\right]$.
To solve above Hamiltonian, the eigenvalues are obtained
\begin{equation}
E_{\pm}=\pm \sqrt{r_{\alpha}^{2}e^{i2\nu_{\alpha}^{(+)}}\sin^{2}\theta_{\alpha}+Z^{2}\cos^{2}\theta_{\alpha}}
\end{equation}
where $r_{\alpha}\equiv\sqrt{r_{\alpha}^{(+)}r_{\alpha}^{(-)}}$, and $\nu_{\alpha}^{(+)}\equiv\frac{\nu_{\alpha,2}+\nu_{\alpha,1}}{2}$.
We define
$\tan\phi_{\alpha}\equiv \frac{r_{\alpha}e^{i\nu_{\alpha}^{(+)}}}{Z}\tan\theta_{\alpha}$,
The corresponding wave functions are given
\begin{equation}
|\psi_{+}\rangle=\left(
\begin{array}{c}
\rho_{\alpha} e^{-i\nu_{\alpha}^{(-)}}\cos\frac{\phi_{\alpha}}{2} \\
\sin\frac{\phi_{\alpha}}{2}
\end{array}
\right);\
|\psi_{-}\rangle=\left(
\begin{array}{c}
-\rho_{\alpha} e^{-i\nu_{\alpha}^{(-)}}\sin\frac{\phi_{\alpha}}{2} \\
\cos\frac{\phi_{\alpha}}{2}
\end{array}
\right)
\label{wf1}
\end{equation}
where $\nu_{\alpha}^{(-)}\equiv\frac{\nu_{\alpha,2}-\nu_{\alpha,1}}{2}$ and $\rho_{\alpha}\equiv\sqrt{\frac{r_{\alpha}^{(+)}}{r_{\alpha}^{(-)}}}$.
Similarly the wave functions in the dual space are
\begin{equation}
|\Lambda_{+}\rangle=\left(
\begin{array}{c}
\frac{e^{-i\nu_{\alpha}^{(-)}}}{\rho_{\alpha}}\cos^{*}\frac{\phi_{\alpha}}{2} \\
\sin^{*}\frac{\phi_{\alpha}}{2}
\end{array}
\right);\
|\Lambda_{-}\rangle=\left(
\begin{array}{c}
-\frac{e^{-i\nu_{\alpha}^{(-)}}}{\rho_{\alpha}}\sin^{*}\frac{\phi_{\alpha}}{2} \\
\cos^{*}\frac{\phi_{\alpha}}{2}
\end{array}
\right),
\label{wf2}
\end{equation}
Similarly, the non-Abelian Berry connection may be given by
\begin{eqnarray}
A & = & i\Lambda^{\dagger}d \Psi \\\nonumber
&=&
i \left(\begin{array}{cc}
\langle\lambda_{+}|d\psi_{+}\rangle & \langle\lambda_{+}|d\psi_{-}\rangle \\
\langle\lambda_{-}|d\psi_{+}\rangle & \langle\lambda_{-}|d\psi_{-}\rangle \\
\end{array}
\right)\\\nonumber
&\equiv &
i \left(\begin{array}{cc}
A^{+}& A^{+,-}\\
A^{-,+} & A^{-}
\end{array}
\right),
\end{eqnarray}
where
\begin{eqnarray}
A^{+}&=&i\langle\Lambda_{+}|d|\psi_{+}\rangle
=\frac{i}{\rho_{\alpha}}\cos^{2}\frac{\phi_{\alpha}}{2}d\rho_{\alpha}+
\cos^{2}\frac{\phi_{\alpha}}{2}d\nu_{\alpha}^{(-)};\\
A^{-}&=&i\langle\Lambda_{-}|d|\psi_{-}\rangle
=\frac{i}{\rho_{\alpha}}\sin^{2}\frac{\phi_{\alpha}}{2}d\rho_{\alpha}+
\sin^{2}\frac{\phi_{\alpha}}{2}d\nu_{\alpha}^{(-)}.
\end{eqnarray}
It can be seen that $\rho_{\alpha}$,
$\phi_{\alpha}$ and $\nu_{\alpha}^{(-)}$ vary periodically with $\varphi_{\alpha}$ and $\theta_{\alpha}$,
which are related to the evolution parameters $\alpha$. Thus, similarly, the global Berry phase can be obtained
\begin{equation}
Q=\frac{1}{2\pi}\oint Tr A=\frac{i}{2\pi}\oint d\ln\rho_{\alpha}+\frac{1}{2\pi}\oint d\nu_{\alpha}^{(-)}
\label{Q2}
\end{equation}
Notice that $\rho_{\alpha}$ is a periodic function of $\alpha$, in a periodicity $\oint d\ln\rho_{\alpha}=\ln\rho_{\alpha}|_{\varphi=0}^{\varphi=2\pi}=0$. Integrating the second term in Eq. (\ref{Q2}), we can obtain:

1. $Q=1$ for $(\Delta_{x}^{2}-h_x^2)(\Delta_{y}^{2}-h_y^2)>0$;

2. $Q=0$ for $(\Delta_{x}^{2}-h_x^2)(\Delta_{y}^{2}-h_y^2)<0$;

3. $Q=1$, for $\Delta_{x}=\Delta_{y}=\Delta_{z}=0$, the system reduces a general two-level Hermitian system.

It can be seen that the global Berry phase $Q$ equal $1$ or $0$ which depends only on $\Delta_{x,y}$ at $\pm h_{x,y}$, but is independent of the values of $\Delta_{x}$ and $\Delta_{y}$ beyond $\pm h_{x,y}$. However, $\Delta_{x,y}=\pm h_{x,y}$ is a singularity and $Q$ has no definition at these points. Therefore, $Q$ can be regarded as a topological index identifying the topological invariance. Thus, when the parameters $\Delta_{x}$ and $\Delta_{y}$ vary crossing $\pm h_{x,y}$, it implies that the topological phase transition occurs.

\subsection{Bipartite dissipative model}
To clearly see the topological invariance and the meaning of complex Berry phase of non-Hermitian systems,
we consider a bipartite one-dimensional (1D) lattice model with dissipation on one of sublattices, which can be realized by the double quantum dots.\cite{Rudner}
An electron can hop between sites and initially localized on any non-decaying site.
The Hamiltonian of this system can be written as\cite{Rudner}
\begin{equation}
H=\sum_{m}\left[\varepsilon _{A} c^{\dagger}_{m}c_{m}
+\varepsilon_{B}d^{\dagger}_{m}d_{m}
+v(c^{\dagger}_{m}d_{m}+d^{\dagger}_{m}c_{m})
+v'(c^{\dagger}_{m}d_{m+1}+d^{\dagger}_{m+1}c_{m})\right]
\label{H}
\end{equation}
where $\varepsilon _{A}$ is the on-site energy of $A$ sites, while
$\varepsilon_{B}=\varepsilon _{A}-i2\Gamma$ is the on-site energy of $B$ sites and the imaginary part
describes the dissipation. The $v$ and $v'$ are the transition amplitudes of intracell and intercell hopping processes respectively. They are independent of the site index $m$, namely the translation symmetry preserves.
Thus, using the Fourier transformation, we can rewrite the Hamiltonian in Eq.(\ref{H}) in the reciprocal space,
\begin{equation}
H=\sum_{k}(c^{\dagger}_{k},d^{\dagger}_{k})\left[
\begin{array}{cc}
\varepsilon _{A} & v_{k} \\
v_{k}^{\ast } & \varepsilon_{B}%
\end{array}
\right]
\left(
\begin{array}{c}
c_{k} \\
d_{k}
\end{array}
\right)
\label{Hk}
\end{equation}%
where $v_{k}=v+v'e^{ik}$. It can be seen that this Hamiltonian is non-Hermitian,
but has different mathematical structure from the first model in Eq.(\ref{Hn1}).
Solving the eigen equation in Eq.(\ref{Hk}) we can obtain the eigenvalues
\begin{equation}
E_{k,\pm}=\varepsilon _{A}-i\Gamma \pm\sqrt{|v_{k}|^{2}-\Gamma^{2}},
\end{equation}%
where $s$ labels pseudospin $\pm1$. The corresponding eigenfunctions are
\begin{equation}
|\Psi _{+}\rangle =\left(
\begin{array}{c}
\frac{v_{k}}{|v_{k}|}\cos\frac{\phi}{2} \\
\sin\frac{\phi}{2}
\end{array}
\right);\
|\Psi _{-}\rangle =\left(
\begin{array}{c}
-\frac{v_{k}}{|v_{k}|}\sin\frac{\phi}{2} \\
\cos\frac{\phi}{2}
\end{array}
\right)
\label{ps}
\end{equation}%
and their dual wave functions in the dual space are
\begin{equation}
|\Lambda _{+}\rangle =\left(
\begin{array}{c}
\frac{v_{k}}{|v_{k}|}\cos^{*}\frac{\phi}{2} \\
\sin^{*}\frac{\phi}{2}
\end{array}
\right);\
|\Lambda_{-}\rangle =\left(
\begin{array}{c}
-\frac{v_{k}}{|v_{k}|}\sin^{*}\frac{\phi}{2} \\
\cos^{*}\frac{\phi}{2}
\end{array}
\right)
\label{lam}
\end{equation}%
where $\tan\phi=\frac{|v_{k}|}{i\Gamma}$, and $\phi$ is a complex variable, and the $k$ is the wave vector within the Brillouin zone $k\in (-\pi,\pi )$.
Similarly, the non-Abelian Berry potential of the system can be given by
\begin{eqnarray}
A &=& i\Lambda^{\dagger}d \Psi \\\nonumber
&=& \frac{1}{2}(\sigma_{0}+\sigma _{z}\cos \phi -\sigma _{x}\sin\phi)\partial _{k}\theta _{k}dk
-\frac{i}{2}\sigma _{y}\partial_{k}\phi dk
\label{GP}
\end{eqnarray}
where $\sigma _{0}=\mathbf{1}_{2\times 2}$; $\sigma _{x,y,z}$ are the Pauli matrices, and
$\theta _{k}$ is defined by $v_{k}=|v_{k}|e^{-i\theta_{k}}$, and $\cos\phi\equiv\frac{i\Gamma}{\sqrt{|v_{k}|^{2}-\Gamma^{2}}}$.
Since the wave function is periodic for $k$, the wave vector $k$ can be regarded as a parameter inducing the Berry phase.\cite{Bohm} The complex geometric phases in two bands can be expressed as
\begin{equation}
\gamma^{G}_{\pm}=\oint_{C}A^{\pm}=\frac{1}{2}\oint(1\pm\cos\phi)d\theta_{k}
\label{GB}
\end{equation}
Notice that the second terms for different states may be canceled, the global Berry phase becomes
\begin{equation}
Q= \frac{1}{2\pi}\oint TrA=\frac{1}{2\pi}\oint d\theta_{k}
\label{T2}
\end{equation}
Let $q=v'/v$ be the ratio of the hopping amplitudes between $A$ and $B$ sites, for $q>1$, as $k$ varies from $-\pi$ to $\pi$, the closed integral path goes around the zero point, which corresponds to the winding number, namely $Q=1$.
for $0<q<1$, the integral closed path does not go around the zero point, it implies $Q=0$. The $Q$ is the winding number of the relative phase between components of the Bloch wave function, which identifies the topological invariance of this quantum dissipative system. This topological invariance identified by $Q$ is robust against the $q$ local variation because $Q=0$ or $1$ depends only on $q<1$ or $q>1$ instead of the $q$ values. This implies a topological invariance and $Q$ can be regarded as a topological index. The TPT occurs at $q=1$ in the parameter space. Namely, this non-Hermitian system has two states that has the topological invariance distinguished by $0<q<1$ and $q>1$ in the parameter space.

On the other hand, the Berry phase in the ground state changed abruptly implies quantum phase transition (QPT) for Hermitian systems.\cite{Carollo} For non-Hermitian systems the complex Berry phase in a quantum state changed abruptly implies also QPT. Suppose that electrons occupy the lower energy band $E_{k,-}$, namely, the half-filling case, the $|\Psi_{-}\rangle$ is the ground state. Thus, the complex Berry phase $\gamma^{G}_{-}$ changed abruptly implies QPT.

In the integration of Eq.(\ref{GB}), for $q<1$, the closed integral path does not go around the zero point such that the first term of the closed path integration equals zero, and for $q>1$, the integral closed path goes around the zero point such that the first term of the integration equals 1. Thus, we have
$\frac{1}{2}\oint d\theta_{k}=\pi\Theta(q-1)$, where $\Theta$ is the step function.
For the second term, notice that $\frac{d\theta}{dk}=\frac{q(q+\cos k)}{1+q^{2}+2q\cos k}=
\frac{1}{2}\left(1+\frac{q^{2}-1}{1+q^{2}+2q\cos k}\right)$,
the second term in Eq. (\ref{GB}) can be written
\begin{eqnarray}
\oint \cos\phi d\theta
&=& \frac{i\eta}{2}\int_{-\pi}^{\pi}\frac{dk}{\sqrt{1+q^{2}+2q\cos k-\eta^{2}}}\\\nonumber
&+& \frac{i\eta(q^{2}-1)}{2}\int_{-\pi}^{\pi}\frac{dk}{(1+q^{2}+2q\cos k)\sqrt{1+q^{2}+2q\cos k-\eta^{2}}}\\\nonumber
&=& \frac{i\eta}{\sqrt{(1+q^{2})-\eta^{2}}}\int_{0}^{\pi/2}\frac{dk}{\sqrt{1-\frac{4q}{(1+q)^{2}-\eta^{2}}\sin ^{2}k}}\\\nonumber
&+& \frac{i\eta(q-1)}{(q+1)\sqrt{(1+q^{2})-\eta^{2}}}\int_{0}^{\pi/2}\frac{dk}{\left[1-\frac{4q}{(1+q)^{2}}\sin ^{2}k\right]
\sqrt{1-\frac{4q}{(1+q)^{2}-\eta^{2}}\sin ^{2}k}}\\\nonumber
\label{int}
\end{eqnarray}
where $\eta=\frac{\Gamma}{v}$. Let $x=\frac{4q}{(q+1)^{2}}$, $y=\frac{4q}{(q+1)^{2}-\eta^{2}}$,
above integration can be rewritten as the Elliptic integral and the complex Berry phase $\gamma^{G}_{\mu}$ can be obtained
\begin{equation}
\gamma^{G}_{\pm}=\pi\Theta(q-1)\pm i\frac{\eta}{2}\sqrt{\frac{y}{q}}
\left(K(y)+\frac{q-1}{q+1}\Pi(x,y)\right)
\label{gamma}
\end{equation}
where $K(y)=\int_{0}^{\pi/2}\frac{dk}{\sqrt{1-y\sin^{2}k}}$ and $\Pi(x,y)=\int_{0}^{\pi/2}\frac{dk}{(1-x\sin^{2}k)\sqrt{1-y\sin^{2}k}}$ are the first and
third kinds of the complete Elliptic integral, respectively.

The analytic properties of $\gamma^{G}_{\pm}$ in Eq. (\ref{gamma}) reveals the relationship between the complex Berry phase, QPT and TPT for non-Hermitian systems.
We plot the $\gamma^{G}_{\pm}$ in the parameter space $(q,\eta)$ in Figs. 1 and 2 for two pseudospin states respectively.

\begin{figure}
\centering
\resizebox{0.6\hsize}{!}{\includegraphics*{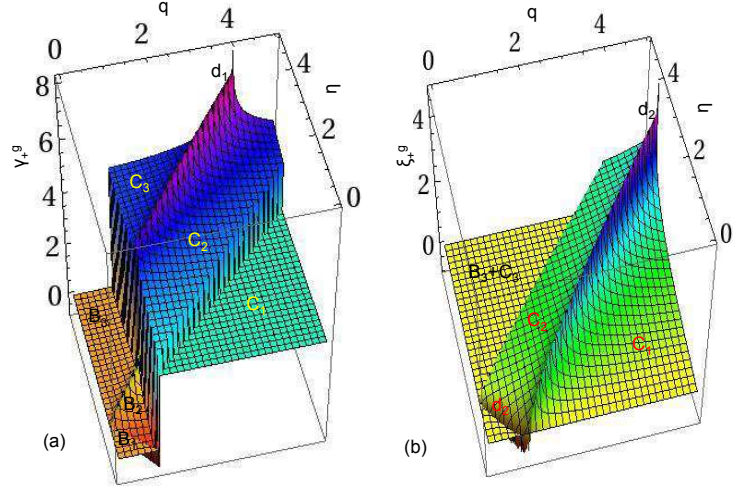}}
\caption{
(Color online)(a)The geometric phases in the up-pseudospin state in the parameter space $(q,\eta)$. The TPT occurs at $q=1$, where $\gamma^{g}_{+}$ has a $\pi$ abrupt change. The states in $0<q<1$ and $q>1$ are two topological phases.   $\gamma^{g}_{+}$ diverged at the line $d_1$, $\eta=q+1$ implies quantum phase transition.
(b) The geometric dissipation $\xi^{g}_{+}$ in the parameter space. $\xi^{g}_{+}=0$ in $B_3$ and $C_3$ means nonexistence dissipative effect from the gauge potential. $\xi^{g}_{+}$ diverges at the line $d_{2}$.}
\label{fig1}
\end{figure}

\begin{figure}
\centering
\resizebox{0.6\hsize}{!}{\includegraphics*{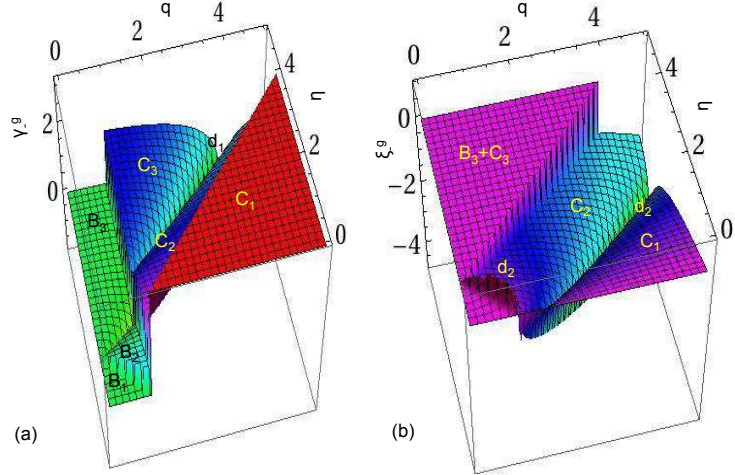}}
\caption{
(Color online)(a)The geometric phases in the down-pseudospin state in the parameter space. Similarly, the TPT occurs at $q=1$, where $\gamma^{g}_{-}$ has a $\pi$ abrupt change.
(b) The geometric dissipation $\xi^{g}_{-}$ in the parameter space. $\xi^{g}_{-}=0$ in $B_3$ and $C_3$ and $\xi^{g}_{-}\ne 0$ in other regions. $\xi^{g}_{-}$ also diverges at the line $d_{2}$.}
\label{fig2}
\end{figure}

From the topological point of views, the parameter space is divided into two parts, $0<q<1$, and $q>1$, where correspond to topological phase (TP) I $Q=0$ and II $Q=1$ respectively, and the Berry phase $\gamma^{g}_{\pm}$ have a step $\pi$ at $q=1$. The abrupt change of $\gamma^{G}_{\pm}$ in the parameter space implies QPT.
In quantum evolution, for Hermitian systems the system properties is sensitive near the exceptional points, where the energy levels are avoided or diabolic crossings with the parameter varying near these points.
For non-Hermitian systems, the eigen energy is complex. When the real part of the eigen energy shown an avoided crossing and the imaginary part a true crossing at the exceptional points is called as the type I exceptional point, and reverse is called as the type II exceptional points.\cite{Keck} We find three basic properties on the exceptional points for this model.

{\bf Claim 3.1}:  In the region $|q-1|\leq\eta\leq q+1$ ($B_{2}+C_{2}$ in Fig. 1 and 2) there exists some $k$ such that the complex energy level is true crossing, $E_{+}=E_{-}$, namely gapless.\\
{\bf Proof}: $E_{+}=E_{-}$ leads to $1+q^{2}+2q\cos k=\eta^{2}$. Namely, $1+q^{2}\pm 2q=\eta^{2}_{max/min}$. It implies that there exists some $k$ satisfying $1+q^{2}+2q\cos k=\eta^{2}$ for $\eta$ within $|q-1|\leq\eta\leq q+1$.

{\bf Claim 3.2}:  In the region $\eta\leq|1-q|$ ($C_{1}$ in Fig. 1 and 2) there exists some $k$ such that the energy-level crossing is type I, $\Im( E_{+})=\Im(E_{-})$, but $\Re(E_{+})\neq \Re(E_{-})$ and implies there exists a gap.\\
{\bf Proof}: For $\eta\leq|1-q|$ $\sqrt{|v_{k}|^{2}-\Gamma^{2}}\in R$ such that $\Re(E_{+})\neq \Re(E_{-})$ and $\Im(E_{+})=\Im(E_{-})$.

{\bf Claim 3.3}:  In the region $\eta\geq 1+q$ ($B_{3}+C_{3}$ in Fig. 1 and 2) there exists some $k$ such that the energy-level crossing is type II, $\Re(E_{+})= \Re(E_{-})$, but $\Im(E_{+})\neq \Im(E_{-})$.\\
{\bf Proof}: For $\eta\geq 1+q$, $\sqrt{|v_{k}|^{2}-\Gamma^{2}}$ is pure imaginary such that $\Re(E_{+})=\Re(E_{-})$ and $\Im(E_{+})\neq \Im(E_{-})$.

The regions in above three claims are sufficient, but not necessary for the conclusion. The exceptional point properties exhibit different energy band structures and the dissipative properties.

The Berry phases $\gamma^{g}_{\pm}$ diverge logarithmically at the line, $\eta=q+1$ labeled by $d_1$ in Figs.1 and 2 (a), but $\gamma^{g}_{\pm}$ jump finitely at the other line, $\eta=|q-1|$ labeled by $d_2$ in Figs. 1 and 2 (b). Similarly, $\xi^{g}_{\pm}$ diverge logarithmically at $\eta=|q-1|$ and jump finitely at $\eta=q+1$.
The $\gamma^{g}_{\pm}$ and $\xi^{g}_{\pm}$ diverged at these lines imply quantum phase transition.\cite{Nesterov}
The $\gamma^{g}_{\pm}$ diverged at the critical line $d_{1}$, $\eta=q+1$, corresponds to resonance by interference of Berry phase and the $\xi^{g}_{\pm}$ diverged at the line $d_{2}$ implies the internal transition of two states by diffusive effects. In other words, the Berry phase $\gamma^{g}_{\pm}$ and its dissipative effect $\xi^{g}_{\pm}$ diverged at the state transition lines correspond to the change of the energy band structure, such as when the parameter changes from $\eta\le|q-1|$ to $\eta\ge|q-1|$ that corresponds the energy band structure changing from gapless to gap, which can be seen from above three claims.

The physical difference between TP I and II is that the system nonexist and exist charge transfer between the system and environment.\cite{Rudner} The bipartite dissipative model is equivalent to the 1D non-Hermitian quantum walk model,\cite{Rudner} Rudner and Levitov define the average displacement of particles as $\langle\Delta m\rangle=\sum_{m}mP_{m}$ to describe the particle decay, where $P_{m}$ measures the decay probability distribution.\cite{Rudner}

\section{Discussions}
It should be remarked that the solutions in Eq. (\ref{psi2}) is obtained under the adiabatic approximation. Actually, there are two regimes for the adiabatic theorem on non-Hermitian Hamiltonian.\cite{Nenciu} For weak non-Hermiticity regime, in which the absolute values of the imaginary parts of the eigenvalues are of the same order of magnetude as the slowness parameter, the adiabatic theorem on non-Hermitian Hamiltonian is valid. For the strong non-Hermiticity regime, in which at least some of the eigenvalues have imaginary parts much larger (in absolute value) than the slowness parameter, the adiabatic theorem on non-Hermitian Hamiltonian is not same to that of the Hermitian Hamiltonian. \cite{Nenciu} Nenciu and Rasche generalize the adiabatic theorem for the strong non-Hermiticity regime that prove an adiabatic expansion exists for the evolution restricted to the subspace corresponding to the least dissipative eigenvalues. \cite{Nenciu}
However, the high-order terms do not affect the topological invariance we address here because we may define the Berry connection without the high-order terms even the high-order terms modify the evolution. In fact, even we define the Berry connection with the first-order term,\cite{Nenciu} we can prove that the first-order term has no contribution to the topological invariance for $2\times 2$ non-Hermitian Hamiltonian. Namely,
\begin{equation}
TrA_{1}=\frac{i}{E_{\ell}-E_{j}}\left[ \langle \partial_{k}\lambda_{\ell}|\psi_{j}\rangle
\langle\lambda_{j}|\partial_{k}\psi_{\ell}
-\langle \partial_{k}\lambda_{j}|\psi_{\ell}\rangle\langle\lambda_{\ell}|\partial_{k}\psi_{j} \right]=0
\end{equation}
where $|\lambda_{\ell(j)}\rangle$ and $|\psi_{\ell(j)}\rangle$ are the eigenvectors in Eqs. (\ref{wf1}),  (\ref{wf2}), (\ref{ps}), and (\ref{lam}). It can be seen that the first-order term in the adiabatic approximation has no contribution to the global Berry phase.

In general, the quantum phase transition for a $N-$level system occurs at the energy-level crossing or avoided crossing (so-called the diabolic or exceptional point) with some parameter varying, which is related only to two energy levels and associates with nonanalytic Berry phase, while other energy levels are invariant with parameters. \cite{Nesterov,Carollo,Sachdev} Thus,
the divergence or jump of Berry phase $\gamma^{g}_{-}$ implies quantum phase transition (QPT)\cite{Nesterov}. However, TPT is characterized by the topological index. The state with topological invariance is robust against local perturbations. TPT is independent of the energy-level crossing and avoided crossing,\cite{Hossein} which can be seen from above discussions. We generalize the global Berry phases $Q$ to identify TPT for two kinds of non-Hermitian models. $Q$ as a topological index describes the topological invariance of systems. $Q$ involves all of the energy band structure.\cite{Zhou,Bohm} It implies that the topological properties of systems depend on the whole energy band structure which actually has be seen from topological insulators.\cite{Kane}
Interestingly, Topological invariance and TPT can occur in both Hermitian and non-Hermitian systems. The topological invariance and TPT in the first model originate from non-Hermitian off-diagonal elements, while the topological invariance and TPT in the bipartite dissipative model are induced by Hermitian off-diagonal elements of the Hamiltonian. The non-Hermitian diagonal element $\Gamma$ does not induce the topological invariance and TPT, but induces QPT.

It should be emphasized that the complex Berry phase was introduced for adiabatic evolution in some specific models. \cite{Keck} The global Berry phase has been used as a topological index for the Hermitian Hamiltonian.\cite{Zhou,Maruyama}
Here we find the gauge invariance of the complex Berry phase and global Berry phase for a generic non-Hermitian Hamiltonian and generalize the global Berry phase to identify the topological invariance for non-Hermitian Hamiltonian. Moreover, we use two two-level non-Hermitian models to reveal the relationship between TPT and QPT.

\section{Conclusions}
In summary, we give the basic properties of the complex Berry phase and the global Berry phases $Q$ of non-Hermitian systems and find that the complex Berry phase and the global Berry phases $Q$ are gauge invariant. We generalize the global Berry phase to non-Hermitian systems as a topological index to identify the topological invariance  for two quite general non-Hermitian models. For the bipartite dissipative model, we give the phase diagram of the complex Berry phase in the parameter space, in which the abrupt change of the Berry phase reveals QPT of the system, which corresponds to the exceptional point. Our findings reveal some topological invariance in non-Hermitian systems, and the relationships between the complex Berry phase, topological invariance, QPT, and TPT in non-Hermitian systems.

\begin{acknowledgments}
The authors gratefully acknowledge the financial support of the project
from the Fundamental Research Fund for the Central Universities.
\end{acknowledgments}

\bibliographystyle{plain}
\bibliography{apssamp}

\end{document}